\documentclass[12pt]{article}
\usepackage{amssymb,amsmath,amsfonts,amsthm}
\usepackage{graphicx}
\usepackage{hyperref}
\usepackage{cite}
\usepackage{subfigure}
\usepackage{epsf}

\hoffset= -1.5cm \voffset= -2cm \vspace{2cm}

\newcommand{\be}{\begin{equation}}
\newcommand{\ee}{\end{equation}}
\newcommand{\bea}{\begin{eqnarray}}
\newcommand{\eea}{\end{eqnarray}}

\begin{document}
\title{Quark-hadron duality, axial anomaly and mixing}

\date{}
\author{Yaroslav~Klopot$^a$\footnote{ On leave from  Bogolyubov Institute for Theoretical Physics, Kiev, Ukraine}\; \footnote{{\bf e-mail}: klopot@theor.jinr.ru} ,\,
        Armen~Oganesian$^{a,b}$\footnote{{\bf e-mail}: armen@itep.ru}\;\, and \
        Oleg~Teryaev$^a$\footnote{{\bf e-mail}: teryaev@theor.jinr.ru}}
\maketitle
\begin{center}
{$^{a}$\em Bogoliubov Laboratory of Theoretical Physics,\\ Joint Institute for Nuclear Research,\;\; \\Joliot-Curie 6, Dubna 141980, Russia.\\
 $^{b}$Institute of Theoretical and Experimental Physics,\;\;\\B. Cheremushkinskaya 25, Moscow 117218, Russia}
\end{center}
\vspace{2cm}

\begin{abstract}
Interplay between axial anomaly and quark-hadron duality in the presence of strong mixing is considered. The anomaly sum rule for meson transition form factors based on the dispersive representation of axial anomaly and quark-hadron duality in octet channel  is analyzed. The comparison of this sum rule to the experimental data on $\eta$ and $\eta'$ mesons transition form factors shows that the interval of duality in this channel is rather small, contradicting the usual understanding of quark-hadron duality. The same values of interval of duality are supported by considering the two-point correlator in the local duality limit. This contradiction may be resolved by introducing of some nonperturbative non-OPE correction to the relevant spectral density. The form and value of this correction are discussed.
\end{abstract}
\newpage
\section{Introduction}

The quark hadron-duality and axial anomaly are the fundamental notions of nonperturbative QCD and hadron dynamics. These problems are combined in description of transition form factors of pseudoscalar mesons, which recently has got considerable interest.

We have analysed these form factors in the previous works \cite{Klopot:2010ke,Klopot:2011qq}  using the exact anomaly  sum rule (ASR) \cite{Horejsi:1994aj} based on dispersive derivation  of axial anomaly \cite{Dolgov:1971ri,Horejsi:1985qu}\footnote{for a review, see \cite{Ioffe:2006ww}.}. The foundation for the Brodsky-Lepage interpolation formula for the pion transition form factor  \cite{Brodsky:1981rp,Radyushkin:1995pj} was obtained. Also, it was shown that the  unusual growth of the pion transition form factor at large virtualities of photon \cite{Aubert:2009mc} can indicate a small (non-OPE) correction to the spectral density.

The further investigation of the ASR in the octet channel of axial current was done in \cite{Klopot:2011qq}. The  form factors were expressed in terms of decay constants  leading to a good description of experimental data \cite{Gronberg:1997fj,:2011hk}. At the same time the calculated interval of duality (continuum threshold) in the suggested approach appeared to be rather small: $s_8\leq m^2_{\eta'}$.

This work is devoted to theoretical investigation of this problem, i.e. how one can combine the axial anomaly, local quark-hadron duality and QCD factorization hypothesis in the octet channel in the case of strong mixing.

\section{Anomaly and duality}
Here we briefly remind the main results of \cite{Klopot:2011qq}.
The dispersive approach to axial anomaly leads to a so-called "anomaly sum rule" (ASR) for one of the scalar amplitudes parametrizing the VVA amplitude for one real  and one virtual  photons \cite{Horejsi:1994aj}. In the  case of octet component of axial current it has the form:
\begin{equation}
\label{ASR} \int_{4m^{2}}^{\infty} A_{3a}(t;q^{2},m^{2}) dt =
\frac{1}{2\sqrt{6}\pi}\;,
\end{equation}
where $A_{3a}(t;q^{2},m^{2})$ is an imaginary part of corresponding invariant  amplitude in the decomposition of the three-point correlation function
\be \label{VVA}
T_{\alpha \mu\nu}(k,q)=\int
d^4 x d^4 y e^{(ikx+iqy)} \langle 0|T\{ J_{\alpha5}(0) J_\mu (x),
J_\nu(y) \}|0\rangle \ee
while $k,q$ are momenta of real ($k^2=0$) and vitual ($q^2=-Q^2$) photons correspondingly.
Let us emphasize, that the equation (\ref{ASR}) is an exact relation, i.e it does not have  $\alpha_s$ corrections  to the integral \cite{Adler:1969er} and (as it is expected from 't Hooft's principle) it does not have nonperturbative corrections as well. The exactness of ASR leads to a very special situation, when it is possible to study the lower-lying states contributions on the top of a large  continuum (higher states) contribution\footnote{Usually one has to suppress the higher states (continuum) contributions as it is done in QCD  sum rules method.}. So one can extract the contributions of lower-lying states which are proportional  to photon-meson transition form factors $F_{M\gamma}(Q^2)$ defined as:

\be \int d^{4}x e^{ikx} \langle M(p)|T\{J_\mu (x) J_\nu(0)
\}|0\rangle = \epsilon_{\mu\nu\rho\sigma}k^\rho q^\sigma
F_{M\gamma} (Q^2). \ee

For this purpose one should saturate (\ref{VVA}) with resonances, use local quark-hadron duality, and finally the ASR  (\ref{ASR}) can be written as follows \cite{Klopot:2011qq}:

\begin{equation}\label{ASR8}
 f_\eta^8 F_{\eta \gamma}(Q^2)+f_{\eta'}^8 F_{{\eta'} \gamma}(Q^2)=\frac{1}{2\pi^2\sqrt{6}}\frac{s_8}{Q^2+s_8},
\end{equation}
where the decay (coupling) constants $f^a_M$ are defined as a projection of axial current onto one-meson states $\eta,\eta'$
\be \langle 0|J^{(a)}_{\alpha 5}(0) |M(p)\rangle=
i p_\alpha f^a_M \;, M=\eta,\eta'. \ee
Since the  relation (\ref{ASR8}) is valid for all $Q^2$, the interval of duality $s_8$ can be determined by considering the limit $Q^2\to \infty$ and matching with QCD factorization:

\be \label{s8}
s_8 = 4\pi^2((f_\eta^8)^2+(f_{\eta'}^8)^2+ 2\sqrt{2} [f_\eta^8 f_{\eta}^0+ f_{\eta'}^8 f_{\eta'}^0]).
\ee
Introducing the vectors of decay constants $\mathbf{e_8}=(f_\eta^8, f_{\eta'}^8)$,  $\mathbf{e_0}=(f_\eta^0, f_{\eta'}^0)$ and charge factors $C^{(8)}=\frac{1}{3\sqrt{6}}, C^{(0)}=\frac{2}{3\sqrt{3}}$ one can rewrite (\ref{s8}) in a compact form:

\be \label{s8c} s_8=4\pi^2 (\mathbf{e_8}^2+\frac{C^{(0)}}{C^{(8)}} \mathbf{e_8\cdot e_0}).
\ee

The formula (\ref{ASR8}) can be compared with the interpolation formulas for the $\eta$ and $\eta'$ transition form factors offered in \cite{Feldmann:1998yc}. Using the expressions for $F_{M\gamma}$ from \cite{Feldmann:1998yc}, the relevant combination of transition form factors in our notation can be written as:

\be \label{FK} f_\eta^8 F_{\eta \gamma}(Q^2)+f_{\eta'}^8 F_{{\eta'} \gamma}(Q^2)=\frac{1}{3\sqrt{6}}(\frac{10f_q^2}{Q^2+4\pi^2f_q^2}-\frac{4f_s^2}{Q^2+4\pi^2f_s^2}),
\ee
where $f_q^2=2(f_{\eta}^0)^2+2(f_{\eta'}^0)^2-(f_{\eta}^8)^2-(f_{\eta'}^8)^2,\; f_s^2=2(f_{\eta}^8)^2+2(f_{\eta'}^8)^2-(f_{\eta}^0)^2-(f_{\eta'}^0)^2$.
One can check that (\ref{FK}) and our result (\ref{ASR8}) (with substituted  $s_8$ from (\ref{s8}))  coincide provided $f_q=f_s$. Otherwise they coincide only in the limits $Q^2=0$ and $Q^2 \to \infty$. For the values suggested in \cite{Feldmann:1998yc} $f_q=1.07 f_\pi,\; f_s= 1.34 f_\pi$ the maximal difference between (\ref{ASR8}) and  (\ref{FK}) is about $10\%$ at $Q^2\thicksim 1\;GeV^2$ .

The discrepancy  has the following origin. The duality interval in anomaly-based Eq. (\ref{ASR8}) corresponds to the octet channel. From the other side, one can easily see that the equation (\ref{FK}) corresponds to the (Born level) quark-hadron duality  applied  for ``light'' and ``strange'' channels recently developed in \cite{Melikhov-temp:2011}. At the same time, the application of anomaly sum rule for these channels requires to take into account the extra contributions due to gluonic anomaly  which are absent in deriving (\ref{ASR8}).

In Ref. \cite{Klopot:2011qq} the ASR (\ref{ASR8}) was applied to several mixing schemes and it appeared to be quite consistent with the experimental data on transition form factors of $\eta$ and $\eta'$ mesons. However, the interval of duality $s_8$  found to be rather small in all the considered schemes: $s_8\leq m^2_{\eta'}$.

\section{Duality and mixing}
It is instructive to consider a simpler case of the duality interval obtained from a two-point correlator of axial currents $J_{\alpha 5}^a,J_{\alpha 5}^a$. The interplay of two- and three-point correlators was investigated for the case of isovector current and pion state in \cite{Radyushkin:1995pj} and the duality interval was expressed in terms of pion decay constant $f_\pi=0.13$ $GeV$:

\be s_3^{\pi}=4\pi^2f_\pi^2.
\ee
This expression was obtained from the local duality limit (Borel parameter $M \to \infty$) of QCD sum rule for two-point correlator which was found to be safe in this case.

The QCD sum rule for octet channel is rather similar: the additional prefactor  $exp(-m^2_{\eta}/M^2)$ is close to 1 for all reasonable not small $M$.  Neglecting the possible instanton contributions and $s$-quark mass effects one gets:

\be s_8^\eta=4\pi^2(f_\eta^8)^2.
\ee

Let us stress, that the model "$\eta+continuum$" cannot be applied  to the case of three-point correlator  since  $\eta'$ meson cannot be included into continuum due to its decay into two real photons; so $\eta'$ should be taken into account explicitly \cite{Klopot:2011qq} which results in a model "$\eta+\eta'+continuum$". Applying it for the two-point correlator in the same local duality limit one arrives to:

\be \label{s8-2p} s_8^{\eta+\eta'}=4\pi^2((f_\eta^8)^2+(f_{\eta'}^8)^2)=4\pi^2 \mathbf{e_8}^2.
\ee

It is interesting to compare the duality intervals obtained from the two- and three-point correlators \footnote{Note, that the anomaly itself is related with the limit $M \to \infty$ of the three-point correlation function \cite{Horejsi:1994aj}.} in this region.
As a result, from (\ref{s8c}) and (\ref{s8-2p}) we see that the duality intervals  coincide only if:

\be \mathbf{e_8}\cdot\mathbf{e_0} =0. \label{cond}
\ee
In terms of decay constants matrix
\be \label{2angF}\mathbf{ F}= \left(\begin{array}{cc}  f_\eta^8 & f_{\eta'}^8 \\ f_\eta^0 & f_{\eta'}^0  \end{array}\right)
\ee
and related to it matrix $\mathbf{E}$
\be \mathbf{E} \equiv  \mathbf{FF^T}= \left(\begin{array}{cc}  \mathbf{e_8}^2 & \mathbf{e_8}\cdot\mathbf{e_0} \\ \mathbf{e_8}\cdot\mathbf{e_0} & \mathbf{e_0}^2  \end{array}\right)
\ee
the condition (\ref{cond}) is also possible to rewrite as:
\be \mathbf{E} = diag(\mathbf{e_8}^2,\mathbf{e_0}^2).
\ee

It easy to check, that this condition remains valid even if more than two mixing states are taken into account (see e.g. \cite{Klopot:2009cm} and references therein). In this case the vectors $\mathbf{e_8}, \mathbf{e_0}$ acquire additional component(s) and mixing matrix $\mathbf{F}$ has a rectangular (nonsquare) form. Let us pass to the consideration of particular cases.

i) The condition (\ref{cond})  is clearly satisfied for the simplest one-angle mixing scheme:

\be \label{1ang} \mathbf{F}=diag(f_0, f_8) \mathbf{U}, \;\;\;\; \mathbf{U(\theta)}=\left(\begin{array}{cc}  \cos \theta & -\sin \theta \\
\sin \theta & \cos \theta  \end{array}\right)
\ee.

The interval of duality is:
\be
s_8=4\pi^2f_8^2.
\ee

ii) Moreover, the  condition (\ref{cond}) is satisfied for mixing schemes where the matrix $\mathbf{F}$ can be  parametrized  as a product of a (rectangular) diagonal and a unitary (orthogonal)  matrices (like those discussed in \cite{Klopot:2009cm}).

iii) At the same time, in the class of schemes where the matrix $\mathbf{F}$ has the following form \cite{Leutwyler:1997yr,Feldmann:1998vh,Escribano:2005qq}

\be \label{2ang} \mathbf{F}= \left(\begin{array}{cc}  f_\eta^8 & f_{\eta'}^8 \\ f_\eta^0 & f_{\eta'}^0  \end{array}\right)=
 \left(\begin{array}{cc}  f_8 & 0 \\ 0 & f_0  \end{array}\right) \left(\begin{array}{cc}  \cos\theta_8 & \sin\theta_8 \\ -\sin\theta_0 & \cos\theta_0  \end{array}\right)
\ee
the matrix $\mathbf{E}$ is  diagonal ($\mathbf{e_8}\cdot\mathbf{e_0}=0$) only if $\theta_8=\theta_0$.

The interval of duality is:
\be
s_8=4\pi^2((f_8)^2+(f_0)^2+2\sqrt{2}f_8f_0(\sin\theta_8\cos\theta_0-\cos\theta_8\sin\theta_0))
\ee

Note, that this kind of matrices may appear when one considers the quark basis \cite{Feldmann:1998vh}.
In terms of this basis the matrix of decay constants is written as $\mathbf{F}=\mathbf{U(\alpha)} diag(f_q,f_s) \mathbf{U(\phi)}$, where $\alpha=\arctan(\sqrt{2})$ and $\phi$ is a mixing angle in the quark basis (see Eq. (2.1) in \cite{Feldmann:1998vh}) and $\mathbf{E}=\mathbf{U} diag(f_q^2,f_s^2)\mathbf{ U}^T$ can be diagonal only if $f_q=f_s$ ($SU(3)$ symmetry). For the values  \cite{Feldmann:1998vh} $f_q=1.07 f_\pi, \; f_s= 1.34 f_\pi$  the matrix $\mathbf{E}=\left(\begin{array}{cc}  0.027 & -0.005 \\ -0.005 & 0.023 \end{array}\right) $ is rather close to a diagonal one. 

So we see that  the intervals of duality $s_8$ corresponding to two- and three- point corelators either coincide, or are very close to each other in all considered schemes. At the same time, their values are quite small $s_8\sim 0.3\div 0.6 $ $GeV^2$.

\section{Corrections}

Earlier, in relation with BaBar puzzle, exploring the exactness of ASR, the existence of small nonperturbative correction to continuum in the isovector channel was supposed \cite{Klopot:2010ke}. It seems natural that the  same kind of correction exists in the octet channel also.

Although the experimental data on $Q^2F_{\eta \gamma}$ and $Q^2F_{\eta' \gamma}$ \cite{:2011hk}
do not show a $Log$-like growth, contrary to pion, the octet combination  of them (corresponding to the l.h.s. of (\ref{ASR8})) indeed manifests a slight growth of this type.

Moreover, there is an additional argument in support of existence of such correction: the mixing is taken into account in the l.h.s. of ASR (\ref{ASR8}) (the sum of resonances) so it should be taken into account in the r.h.s. also. Indeed, the one-loop approximation for the spectral density $A_{3a}^{QCD}(s,Q^2)=\frac{1}{2\pi\sqrt{6}}\frac{Q^2}{(Q^2+s)^2}$ does not take into account mixing. So it can be taken into account  via $1/Q^2$ correction to the spectral density $\delta A_{3a}$ of a non-OPE origin.\footnote{Two-loop corrections are  proven to be zero \cite{Jegerlehner:2005fs}.}  Let us remind, that the full integral $\int_{0}^{\infty}( A_{3a}^{QCD}+\delta A_{3a}) ds$ has no corrections, i.e. $\int_{0}^{\infty} \delta A_{3a} ds=0$, though the continuum part may have a correction: \be I_{cont}=I_{cont}^{QCD}+\delta I_{cont},\ee where

\be I_{cont}^{QCD}=\int_{s_8}^{\infty} A_{3a}^{QCD} ds=\frac{1}{2\pi\sqrt{6}}\frac{Q^2}{s_8+Q^2}, \;\;\; \delta I_{cont}=\int_{s_8}^{\infty} \delta A_{3a}ds.\ee
Such correction should be small $\delta I_{cont}/I_{cont}^{QCD}\ll 1$ and usually can be neglected. But because of the exactness of ASR (\ref{ASR}) it results in the relatively large corrections to the lower states.

Note, that $\delta I_{cont}$  must satisfy the following four limits: $\delta I_{cont}=0$ at $Q^2=0$, $Q^2 \to \infty$,  $s_8=0$ and  $s_8 \to \infty$.
Its simplest possible form is:

\be \label{corr}
\delta I_{cont}=-I_{cont}^{QCD}\frac{c s_8}{(Q^2+s_8)}(\ln(Q^2/s_8)+b),
\ee
where $b,c$ are  parameters to be fitted. Then the ASR takes a form:

\begin{eqnarray} \label{fit}
\nonumber f_\eta^8 F_{\eta \gamma}(Q^2)+ f_{\eta'}^8 F_{{\eta'} \gamma}(Q^2) = \frac{1}{\pi}\left(\frac{1}{2\pi\sqrt{6}}-I_{cont}^{QCD}-\delta  I_{cont}\right)= \\\frac{1}{2\pi^2\sqrt{6}}\frac{s_8}{Q^2+s_8}\left(1+\frac{c Q^2}{(Q^2+s_8)}(\ln(Q^2/s_8)+b)\right).
\end{eqnarray}

It is important, that such kind of correction violates the factorization hypothesis and so it does not allow to determine the duality interval $s_8$ from ASR at large $Q^2$ as it was done in previous section.
At the same time, numerical analysis shows that the  relation (\ref{fit}) can be fulfilled for the conventional  estimations of continuum threshold $s_8=1.5-2.5 GeV^2$ \cite{Ioffe:1999uj,Aliev:2003ji}.

In Fig.1 the plot of Eq. (\ref{fit})  for the case of one-angle mixing scheme  is shown ($\theta=-16^o$, and $f_8=0.94 f_\pi$ is determined from ASR at $Q^2=0$). The l.h.s. of (\ref{fit}) is evaluated from the data on $\eta,\eta'$ transition form factors \cite{Gronberg:1997fj,:2011hk}, and fit of (\ref{fit}) is performed for different values of  duality interval: $s_8=0.5$ $GeV^2$ $(b=-2.63, c=0.23)$,   $s_8=1.5$ $GeV^2$ $(b=-7.13, c=0.14)$ and $s_8=2.5$ $GeV^2$ $(b=-7.5, c=0.16)$. So, for the ``usual'' values of continuum threshold $s_8=1.5-2.0$ $GeV^2$ the experiment can be described well, while  for $s_8>2.5$ $GeV^2$  the fit is substantially worse.

\begin{figure}
\includegraphics[width=0.6\textwidth]{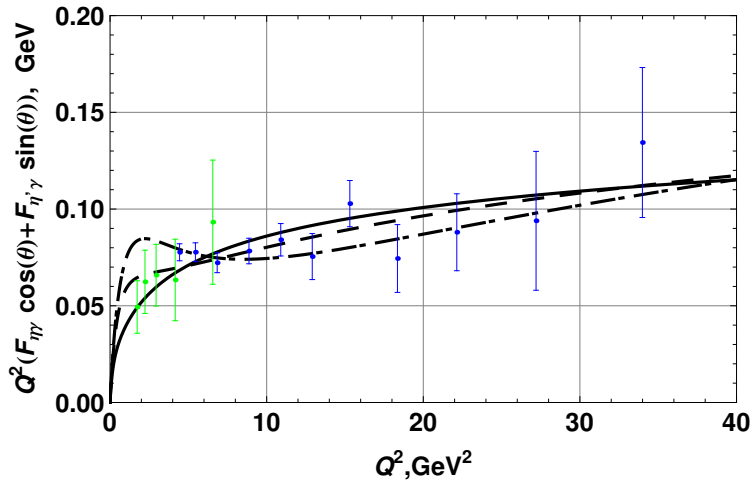}
 \caption{Eq. (\ref{fit})  for the one-angle mixing scheme: $s_8=0.5,\; 1.5, \; 2.5$ $GeV^2$ (solid, dashed, dot-dashed curves respectively).}
\end{figure}

\section{Discussion and conclusions}

The  approach  based on dispersive representation of axial anomaly allows to justify the quark-hadron duality for transition form factors of pseudoscalar mesons. In this paper we concentrated on the analysis of the octet channel  where a strong mixing of $\eta$ and $\eta'$ mesons is manifested. As a result we obtained the  relation between the transition form factors and decay constants. This relation is found to be in a reasonably good agreement with  experimental data for various mixing schemes.

At the same time the following problem emerges:  the duality interval (which follows from anomaly matching with factorization hypothesis) appears to be surprisingly small in all the considered mixing schemes: $ s_8\lesssim 0.7\; GeV^2$. This is dramatically smaller than the conservative value $2-2.5$ $GeV^2$ estimated from the squared mass of the first higher resonance.

The first observation to be made is that this duality interval may be considered as  corresponding to octet channel rather than to a particular state. This means that substantially  heavier resonances ``stranger'' to the given channel (like $\eta'$ to the octet channel) manifest themselves in the duality intervals characteristic of the lighter ``host'' resonances. Let us stress that the situation of mixing of hadrons with substantially different masses is rather uncommon (typically the states with close masses like $\rho$ and $\omega$ mix) and in the case of $\eta$ and $\eta'$ mesons emerges due to anomalies. Indeed, on the one hand, the large mass of $\eta'$ is generated by the gluonic anomaly (the famous $U_{A}(1)$ problem, see \cite{'tHooft:1986nc,Diakonov:1981nv} and references therein). On the other hand, the abelian anomaly at $Q^2=0$ indicates \cite{Klopot:2009cm} a significant mixing.

The appearance of small duality interval in the octet channel is also confirmed by the analysis of correlator of two octet axial currents. The relatively small coupling of $\eta'$ meson to octet channel $f_{\eta'}^8$ results in the small duality interval (see Eq. (\ref{s8-2p})) in concordance with above mentioned picture. It is important, that the duality intervals calculated from the two- and three-point correlation functions appear to be close to each other or even coincide provided the particular constraint (Eq. (\ref{cond})) for mixing scheme holds.

To interpret these results one may say that the quark-hadron duality in the presence of strong quark-hadron mixing is manifested in a very special way in the form of the semi-local duality (being intermediate between local and global ones), when the meson in the alien channel is represented by the unexpectedly small duality interval whose position is not tightly correlated with its mass.

Another way to treat the specifics of quark-hadron duality in the presence of strong mixing is to assume the existence of particular nonperturbative corrections. Such corrections were first  introduced \cite{Klopot:2010ke} as a possible explanation of BABAR puzzle for pion (isovector channel).
It is natural to suppose that such  corrections exist in the octet channel too. Note, the lower states in both isovector and octet channels are Goldstone bosons which can have flat distribution amplitudes (see e.g. \cite{Radyushkin:2009zg,Polyakov:2009je}) resulting in a violation of QCD factorization. As the correction under discussion also leads to violation of QCD factorization, the whole picture is self-consistent.

Moreover, the existence of corrections  in the case of mixed states are supported by the additional  arguments: mixing requires a QCD mechanism responsible for its emergency. These corrections may be represented by some non-OPE contributions, originated possibly from short strings \cite{Chetyrkin:1998yr,Chernodub:2000bk}  or  instantons. Numerical analysis (performed in Section 4) shows that in this case the duality interval can be essentially larger approaching the conservative values $1.5-2 \; GeV^2$.

To summarize, one can consider the situation with a small interval of duality from two points of view. One may  attribute the interval of duality to the channels, rather than to particles themselves, or, otherwise, to preserve the "conservative" values (following from the usual understanding of quark-hadron duality)  one can suppose the existence of non-OPE corrections to continuum contribution.

We thank  B.~L.~Ioffe, P. Kroll, D.I. Melikhov, M.~V.~Polyakov and  A.~V.~Radyushkin for helpful discussions and valuable comments. This  work was supported in part by RFBR (grants 09-02-00732, 09-02-01149, 11-02-01538, 11-02-01454) and by fund from CRDF  Project RUP2-2961-MO-09.

\end{document}